# A Unity of Science, Especially Among Physicists, Is Urgently Needed to End Medicine's Lethal Misdirection :
# Chapter 3. Prevention of the World Avian Flu Pandemic

John T. A. Ely, Ph.D.
Radiation Studies, Box 351310
University of Washington
Seattle WA 98195

In the 1997 Hong Kong bird flu epidemic, the highly pathogenic avian influenza virus exhibited a strain (H5N1) that passed from poultry to people (18 infections and 6 deaths). It is feared[1] that additional mutation may enable transmission of the H5N1 strain directly between humans, leading rapidly to a global pandemic. This strain re-emerged in 2003 in 9 Asian countries with over 120 cases and in excess of 60 deaths to date. It has now spread in Africa, China, the Middle East and Europe. A method for vaccination induction in humans of antibodies and cellular immunity against H5N1 has been tested for use to resist pandemic spread of lethal avian influenza[1]. Now, we show below that even without use of this slow and expensive vaccination method above, the entire human population can, in principle, protect itself (at essentially no cost to the UN) and also simultaneously weaken the avian flu pandemic.

In some publicized locations, many (~50%) of the untreated humans infected with avian flu do not die[2]. These surviving patients are likely unaware of ascorbic acid (AA) or that all humans unknowingly obtain it in food. Their infection is sufficiently mild that the amount of AA they absorb from food is high enough (possibly circa 100 mg/day) so their white cells can protect them against the level of flu virus involved; see food AA content database[3]. The present proliferation of H5N1 infections demands all humans immediately maximize their food AA using the database[3] for guidance. In avian flu patients (or those at risk), survival is a simple matter of whether AA is sufficient that the virus dose is depleted before the AA. Obtaining AA in food is an effective low cost strategy. Humans cannot resist and or recover from any flu (or other infection) without AA (commonly called Vitamin C although it is not a vitamin[4]).

A high sugar diet is undesirable for anyone because it greatly impairs the utilization of AA. In even "modest" blood sugar elevations, glucose molecules so outnumber AA that they competitively inhibit insulin-mediated active transport of AA into cells. This is called the glucose ascorbate antagonism (GAA)[4,5].

In summary: 1) exposure must be minimized to decrease viral load; 2) AA intake must be maximized; and 3) free sugar must be avoided.